\documentclass{aa}
\usepackage{graphicx}
\usepackage{txfonts}
\usepackage{natbib}
\bibpunct{(}{)}{;}{a}{}{,}
\usepackage{lscape}
\DeclareGraphicsExtensions{.pdf,.png,.jpg}
\newcommand{\mic}{\mbox{$\,\mu$m}}

\begin{document}
   \title{Properties of CO$_2$ clathrate hydrates formed in the presence of MgSO$_4$ solutions with implications for icy moons}
    \titlerunning{Properties of CO$_2$ clathrate hydrates in MgSO$_4$ solutions}
    
   \author{E. Safi\inst{1,2}\thanks{email: e.safi@keele.ac.uk}
          \and Stephen P. Thompson\inst{2}
	  \and Aneurin Evans\inst{1}
          \and Sarah J. Day\inst{2}
      \and C. A. Murray\inst{2}
          \and J. E. Parker\inst{2}
      \and A. R. Baker\inst{2}
          \and J. M. Oliveira\inst{1}
          \and J. Th. van Loon\inst{1}
}

   \institute{Astrophysics Group, Lennard-Jones Laboratories, Keele University, Keele, Staffordshire, ST5
    5BG, UK
         \and Diamond Light Source, Harwell Science and Innovation Campus,
	 Didcot, Oxfordshire, OX11 0DE, UK
        }

   \date{Version 25, \today}

% \abstract{}{}{}{}{}
% 5 {} token are mandatory

  \abstract{There is evidence to suggest that clathrate hydrates have a significant
  effect on the surface geology of icy bodies in the Solar System. However the aqueous
  environments believed to be present on these bodies are
  likely to be saline rather than pure water.
  Laboratory work to underpin the properties of clathrate hydrates in
  such environments is generally lacking.}
  {We aim to fill this gap by carrying out a laboratory investigation
  of the physical properties of CO$_2$ clathrate hydrates produced in weak 
  aqueous solutions of MgSO$_4$.}
  {We use in situ synchrotron X-ray powder diffraction to investigate clathrate hydrates formed 
  at high CO$_2$
  pressure in ice that has formed from aqueous solutions of MgSO$_4$ with varying concentrations.
  We measure the thermal expansion, density and dissociation properties of the clathrates under 
  temperature   conditions similar to those on icy Solar System bodies.}
  {We find that the sulphate solution inhibits the formation of clathrates 
  by lowering their dissociation temperatures.
  Hysteresis is found in the thermal expansion coefficients as the 
  clathrates are cooled and heated; we attribute this to the presence
  of the salt in solution. We find the density derived from X-ray powder diffraction measurements
  is temperature and pressure
  dependent. When comparing the density of the CO$_2$ clathrates to that of
  the solution in which they were formed, we conclude   that they should sink
  in the oceans in which they form.
  We also find that the polymorph of ice present at low temperatures is Ih
  rather than the expected Ic, which we tentatively attribute to the presence of
  the MgSO$_4$.}
  {We (1) conclude that the density of the clathrates has implications for their behaviour in 
  satellite oceans as their sinking
  and floating capabilities are temperature and pressure dependent, (2) conclude that the presence
  of MgSO$_4$ inhibits the formation 
  of clathrates and in some cases may even affect their structure and (3) report the dominance of
  Ih throughout the experimental procedure despite 
  Ic being the stable phase at low temperature.}

   \keywords{Methods: laboratory --
     Molecular data --
   Planets and satellites: surfaces --
   Planets and satellites: individual: Europa --
   Planets and satellites: individual: Enceladus
   }

   \maketitle
%
%________________________________________________________________

\section{Introduction}

\label{ChSS}
Clathrate hydrates are formed at high pressures and low temperatures and are
cage-like structures in which water molecules bonded via hydrogen bonds can encase
guest molecules. The conditions on icy Solar System bodies
such as Enceladus, Europa, Mars and comets have long been considered as potential for clathrate
formation \citep{Max2000, Prieto-Ballesteros2005, Marboeuf2010, Bouquet2015}.
Clathrates are leading candidates for the storage of gases such as CH$_4$ and CO$_2$
in the Solar System \citep{Prieto-Ballesteros2005, Mousis2015, Bouquet2015};
therefore understanding the kinetics and
thermodynamics of clathrate hydrates under planetary conditions is important.

The type of guest molecule that can be trapped within a clathrate depends on
the clathrate structure, of which three are currently known: 
sI, sII and sH \citep{Sloan2007}. 
sI clathrates form a cubic structure with space group Pm-3n. They are composed
of two cage types; the smaller 5$^{12}$ (12 pentagonal faces) and the larger
5$^{12}$6$^2$ (12 pentagonal faces and 2 hexagonal faces). sI clathrates are
constructed of two small cages for every six larger ones, and host relatively
large molecules such as CO$_{2}$ and CH$_{4}$. sII also form cubic structures and are composed of
sixteen small 5$^{12}$ cages and eight large 5$^{12}$6$^4$ cages; sII
clathrates typically host smaller molecules such as O$_{2}$ and N$_{2}$.
The least common clathrate hydrate, sH, is composed of one large cage,
three smaller cages and two medium 4$^{3}$5$^6$6$^{3}$ cages. sH clathrates form
hexagonal structures and  
usually require two types of guest species in order to remain stable. Recently a new
structure of clathrate has been proposed \citep{Huang2016}; this type of
clathrate (``structure~III'') is predicted to have a cubic structure and be
composed of two large  8$^6$6$^8$4$^12$ and six small 8$^2$4$^8$ cages.

It has been confirmed that water ice and CO$_{2}$ are present on the
surface of Enceladus \citep{Matson2013}. Among the gases such as CH$_4$
present in the plumes emanating from the satellite's surface \citep{Waite2006}, CO$_{2}$
has poor solubility in water. This
suggests the trapping of gases in the form of clathrate hydrates,
with subsequent release due to their dissociation
\citep{Bouquet2015} could give rise to the plumes. \citet{Fortes2007}
used a clathrate xenolith model to account for the origin of Enceladus' plumes,
suggesting that fluids are able to break through the ice shell, metasomatising
the mantle by the emplacement of clathrates along fractures and grain
boundaries. The clathrates are trapped in the rising cryomagmas as xenoliths,
and are carried upwards where they dissociate, releasing their
enclosed gas and forming the plumes.

The density of clathrates is a significant factor in
determining their fate. If their density is higher than that of
the oceans in which they are formed, they sink to the
ocean floor; if lower, they rise to the ice/ocean interface.
If their destination is the ocean floor then they might
be dissociated by heat produced from hydrothermal activity.
On the other hand if they ascend, then a clathrate layer would be present
at the interface between the ice and ocean surface.

\citet{Bouquet2015} calculated the density of clathrates
assuming fully filled cages and a volatile composition based on
Enceladus' plume; they found densities of 1.04~g~cm$^{-3}$ and
0.97~g~cm$^{-3}$ for sI and sII clathrates respectively. When
comparing these to their computed ocean densities they
deduced that sII clathrates should be buoyant and therefore likely to
ascend. However they were
unable to arrive at a conclusion regarding the sI clathrates,
as there was significant uncertainty regarding the ocean's salinity,
and because the clathrate density was too close to that of the ocean
itself. Clathrate ascension would, however, enable clathrates to play a part
in the formation of the plumes, as their dissociation would
increase pressure conditions at the site of the plume's origin \citep{Bouquet2015}. 

The trapping of gases by clathrates could also have a significant
impact on Enceladus' ocean composition and hence the plumes emitted in the south polar region
\citep{Bouquet2015}.
The enclathration of gases would lower the concentration of volatiles in the ocean
to below that observed in the plumes. This would indicate that any clathrates formed would need to
dissociate in order to replenish the volatile concentration of the plume. If this is
not the case then the gas concentration would need to be restored by an
alternative mechanism, such as hydrothermal activity \citep{Bouquet2015}.

\citet{Prieto-Ballesteros2005} evaluated the stability and
calculated the density of several types of clathrates thought to
be found in the crust and ocean of Europa using thermal models for
the crust. They found SO$_2$, CH$_4$, H$_2$S and CO$_2$
clathrates should all be stable in most regions of the crust. They
deduced that CH$_4$, H$_2$S and CO$_2$ clathrates should float in an
eutectic ocean composition of MgSO$_4$-H$_2$O, but that SO$_2$ clathrates
would sink. However the sinking and floating capabilities of various hydrates
will also likely depend on the salinity of the ocean since this will affect their buoyancy.

\citet{Mousis2013} investigated  clathrates in Lake Vostok (Antarctica) using a 
statistical thermodynamic model.
They assumed temperatures of 276~K and pressures of 35~MPa and
found that Xe, Kr, Ar and CH$_4$ should be depleted in
the lake, while CO$_2$ should be enriched compared to its atmospheric
abundance. They also found that air clathrates should float as they are
less dense than liquid water. However, air clathrates
have not been observed on the surface of the ice above the lake
\citep{Siegert2000}. To account for this \citet{McKay2013}
suggested that large amounts of CO$_2$ are also trapped within the clathrates,
increasing their relative density.

Clathrates have been found to form in the Sea of Okhotsk (Pacific Ocean), and
\citet{Takeya2006} have used X-Ray powder diffraction to study their crystal
structures and thermal properties. They found that four samples from four
different locations each
had sI clathrates encaging CH$_4$ and a small amount of CO$_2$. 
The small amount of encaged CO$_2$
is consistent with \citeauthor{McKay2013}'s suggestion
that a large amount of trapped CO$_2$ is necessary for clathrates
to sink.

\begin{figure}
\begin{center}
\includegraphics[width=9.cm,keepaspectratio]{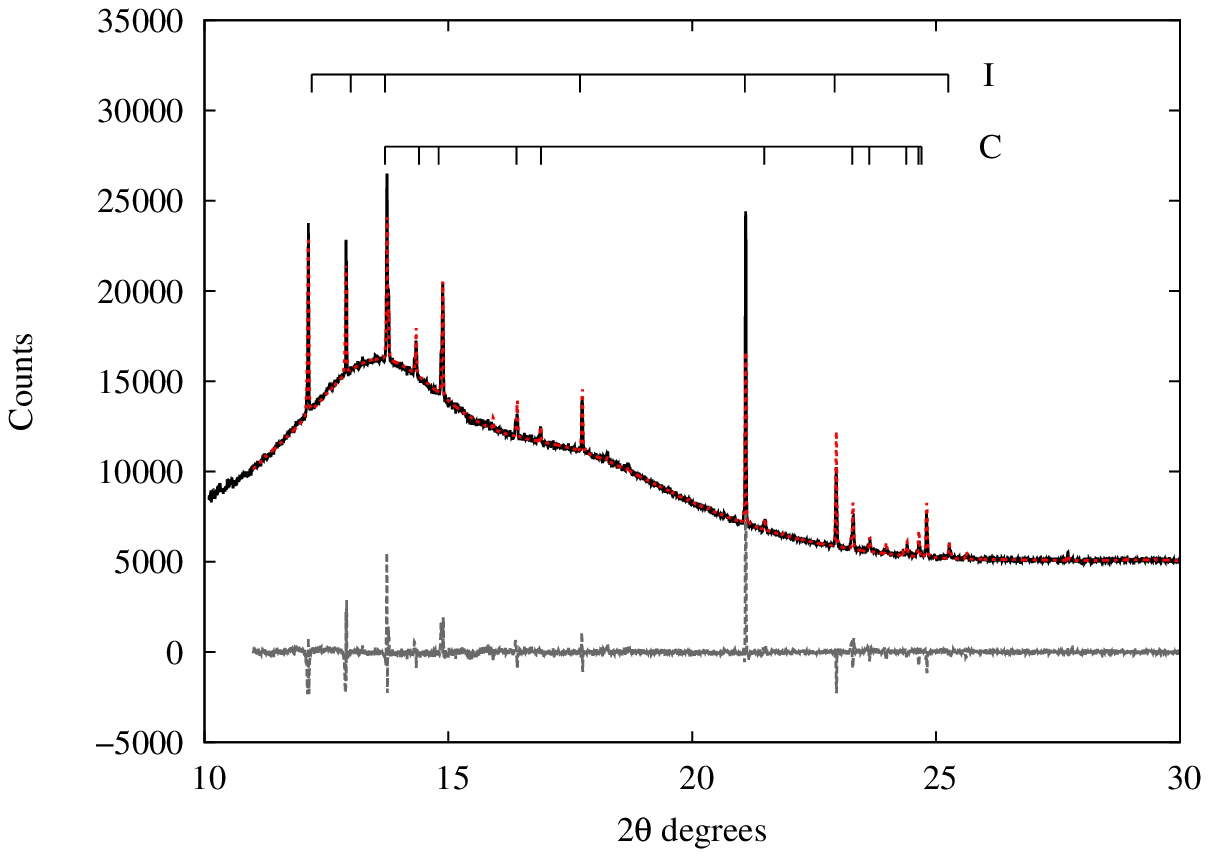}
\includegraphics[width=9.cm,keepaspectratio]{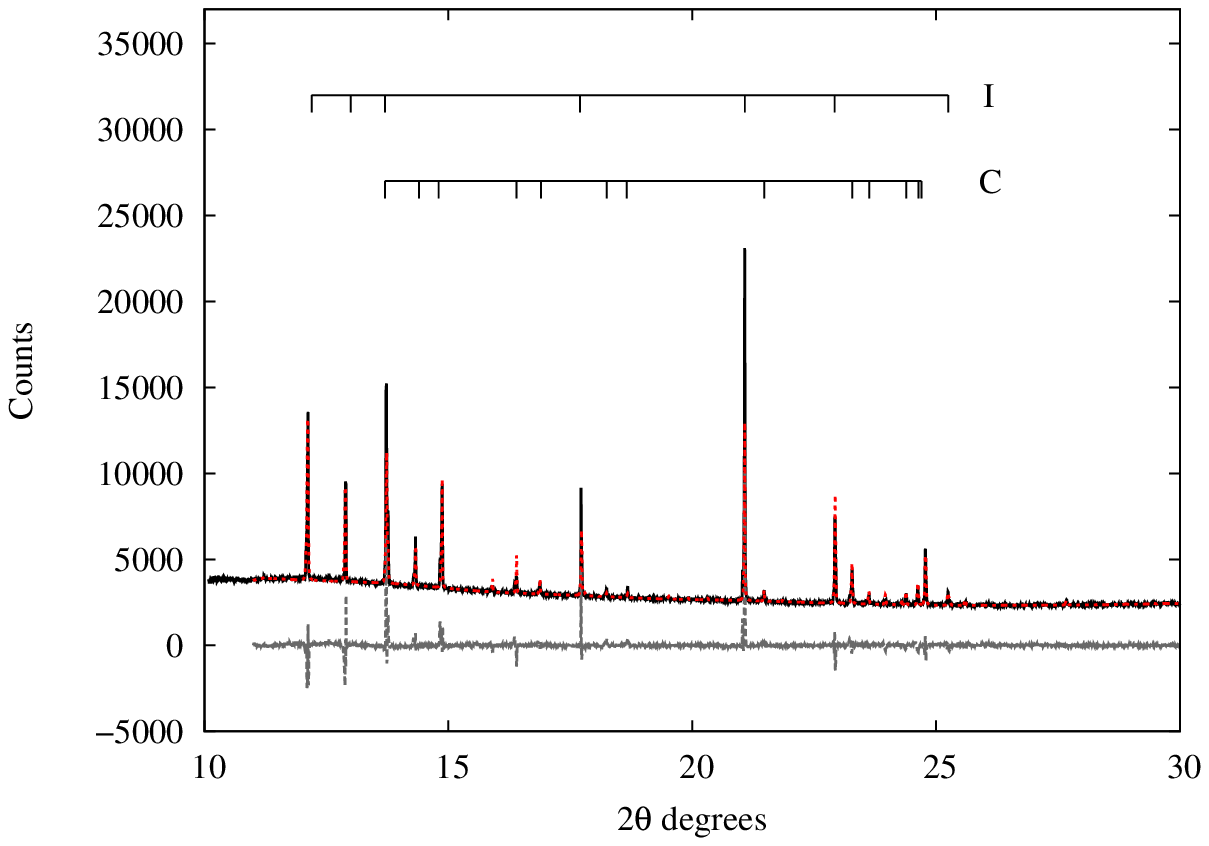}
\includegraphics[width=9.cm,keepaspectratio]{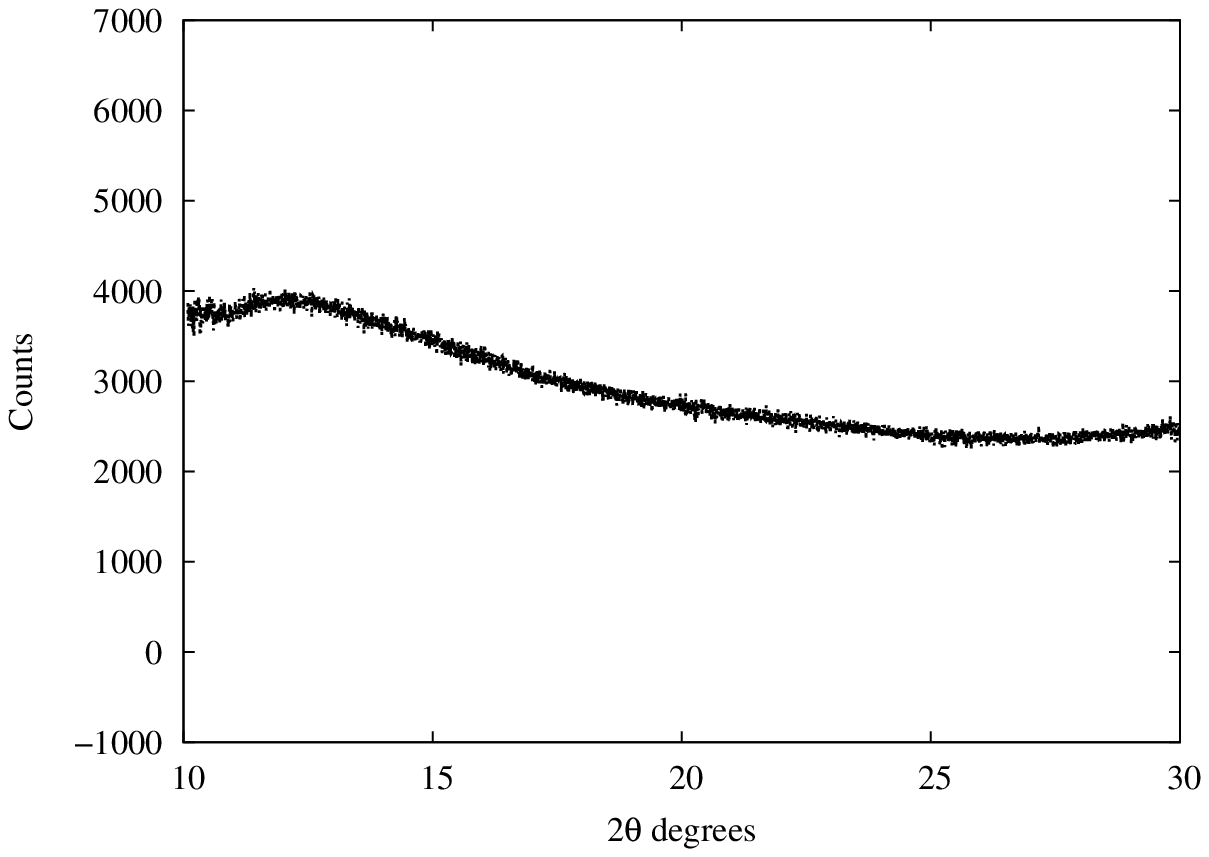}
\caption{Rietveld refinements of the MS10.5 solution
at a 10 bar CO$_2$ pressure experimental data. 
From top-bottom: 90~K, 180~K, 245~K. The experimental data are shown
in black, the calculated fit in red, and the residuals in grey below.
All prominent peaks are labelled where C = clathrate peaks and I = hexagonal
ice peaks.
The larger residuals for some of the ice peaks are due to
poor powder averaging due to the way the ice freezes inside the cell
(preferred orientation) and the restricted cell rocking angle used
to compensate for this during measurement.
\label{pawely}}
\end{center}
\end{figure}

It is likely that the type of ice present during the
formation of clathrates also has an effect on their dissociation.
Using neutron diffraction \citet{Falenty2015} studied the
dissociation of CO$_2$ clathrates in pure water ice between
170--190~K, with special attention to the polymorph of
ice formed. They found that below 160~K cubic ice (Ic)
was the more stable phase, while between 160--190~K Ic transforms to
the more thermodynamically stable hexagonal ice (Ih). \citet{Falenty2015} concluded that,
due to Ic forming with smaller crystallite sizes compared to Ih,
it could provide an additional pathway for the escaped gas molecules originating from clathrate structures,
therefore supporting their dissociation.

An important factor when considering clathrate
dissociation is ocean salinity. It is well known that saline solutions depress the
freezing point of water \citep[e.g.][]{Duan2006}, suggesting
that the temperatures at which clathrates form and dissociate within
the oceans on planets and satellites will also be lower.
\citet{Miller1961} showed how clathrate dissociation is affected by temperature
and pressure conditions in pure water. However, when comparing the
results obtained by Miller with the more recent theoretical
results obtained by \citet{Bouquet2015} for saline solutions,
there is a significant difference, suggesting that increased salinity 
may indeed lower the temperature at which clathrates are able to form.

In this paper we use synchrotron X-ray powder diffraction (SXRPD)
to investigate the thermal and
physical properties of CO$_2$ clathrate
hydrates produced from weak aqueous solutions of MgSO$_4$.
We replicate possible
thermal variations due to seasonal and tidal changes, ocean
depth and salt concentration and observe the formation and
dissociation conditions of CO$_2$ clathrate hydrates.
The SXRPD provides information about the temperature-dependence of clathrate
densities and hence their ability to rise or sink in the oceans in which they are
formed. We also investigate clathrate dissociation kinetics and
the influence of the different polymorphs of ice. 

\begin{table*}
\centering
\caption{Concentration, temperature range, pressure, density of salt solutions.
For comparison, the wt\% salinity of Enceladus' ocean is estimated to be in the range
$2-10$~g salt per kg H$_2$O \citep{Zolotov2007}, that of
Europa is estimated to lie between 1.1 and 96.8~g salt (MgSO$_4$)
per kg H$_2$O \citep{Hand2007}.
\label{salts}}
\begin{tabular}{ccccc}
\hline
Concentration      & wt\%  & Temperature         & Pressure    &  Density     \\ [1ex]
%                    &           &                     &             & \\ [1ex]
 g MS7/kg H$_2$O & MgSO$_4$/kg H$_2$O           & range ($\pm 5$) (K)  &  (bar)      &  (g~cm$^{-3}$)\\ [1ex] \hline
    20             &    10.5 (MS10.5)   &   90.1 -- 225.02    &  $5\pm 0.01$          &    $1.016\pm0.001$\\
   20              &   10.5 (MS10.5)   & 90.01 -- 240        & $10\pm 0.01$          &    $1.016\pm0.001$\\
   5               &    3.1 (MS3.1)     &  89.96 -- 245.04    &  $20\pm 0.01$         &    $1.003\pm0.001$\\[1ex]
\hline
\end{tabular}
\end{table*}

\section{Experimental work}
\label{experimental}
In this work we use an epsomite (MgSO$_{4}\cdot$7H$_2$O)
salt solution to form the ice and CO$_2$ clathrate system. 
The concentrations, and the temperature and pressure ranges used,
are summarised in Table~\ref{salts}, in which
the concentration of MgSO$_{4}\cdot$7H$_2$O has been converted
to concentration of MgSO$_4$ per kg~H$_2$O, allowing for the
contribution that the waters of hydration make to the achieved concentrations. 
In the following we refer to 20g MgSO$_{4}\cdot$7H$_2$O/1kg H$_2$O
as MS10.5, and 5g MgSO$_{4}\cdot$7H$_2$O/1kg H$_2$O as MS3.1.
The salt concentrations are 
similar to those of Enceladus \citep[whose salinity
is estimated to lie in the range 2--10~g/kg H$_2$O;][]{Zolotov2007} 
and of Europa \citep[MgSO$_4$ concentration
estimated to be between 1--100~g/kg H$_2$O;][]{Hand2007}.

The temperature ($T$) range we cover is from $\sim90$~K to $\sim~240$~K,
and the bulk of our measurements were carried out at pressures ($P$) of 5, 10 and 20~bar. 
The range of $T$ is somewhat below that estimated for the sub-surface oceans of (for
example) Europa and Enceladus \citep[see e.g.][]{melosh,Bouquet2015}, and is more
representative of these satellites' surfaces.
The pressures we used in this work were necessarily optimised to give a reasonable 
conversion rate to clathrate with the facility, and within the time, available.
In planetary environments the pressures in sub-surface
oceans depend on the depth of the overlying ice sheet, but are typically hundreds of bar
\citep[see e.g.][]{melosh}, although the pressures in Enceladus' sub-surface ocean may
be as low as a few 10s of bars \citep{Matson2012,Bouquet2015}.

SXRPD data were collected using beamline I11 at the
Diamond Light Source  \citep{Thompson2009} 
during twelve 8-hour shifts.
The X-ray wavelength was 0.826220\,\AA, 
calibrated against NIST SRM640c standard Si powder; the beam size at the sample 
was 2.5~mm (horizontal) $\times$ 0.8~mm (vertical).
The high pressure gas cell and the procedure used to form clathrates are
described by \citet{Day2015}.
A 0.8~mm diameter single-crystal sapphire tube is filled with solution and sealed 
into the gas cell. This is then mounted onto the central circle of the beamline's
concentric three circle diffractometer, and cooled using a liquid nitrogen
Oxford Cryosystems 700+ cryostream. The latter has
temperature stability $\pm0.1$~K and a ramp rate of 360~K/h.

Once frozen at $\sim$240 K, CO$_2$ gas is admitted to the cell at the chosen pressure
and a fast position sensitive detector \citep{Thompson2011} is used to collect
in situ powder diffraction data as the temperature is slowly raised. During this
time ice and clathrate formation is simultaneously observed.
We continue to increase the temperature
until both the clathrate and ice are lost, whereupon the temperature ramp is reversed
and the cell is cooled once more. Depending on pressure and solution composition,
either pure-phase clathrates or an ice-clathrate mix is formed. Using this ``second cycle''
technique provides increased clathrate formation \citep[see discussion in][]{Day2015}.
For the present work we then cycled the temperature between 250~K and 90~K
using the Cryostream to replicate diurnal and tidal variations with applied CO$_{2}$
pressures between 5--20~bar. Dissociation temperatures
and pressures were determined by holding the sample at constant
pressure and gradually increasing the temperature in 5~K temperature
steps until there were no peaks discernible in the X-ray diffraction pattern.

Each SXRPD data-collection
cycle, including the time allowed for the sample to come to
temperature equilibrium, took approximately 20~minutes.
Once data collection was completed the temperature was changed to the new setting
and data collection repeated.

The SXRPD patterns were analysed via Rietveld structure refinement, using TOPAS
refinement software \citep{Coelho2007} and previously published clathrate atom positions and lattice
parameters \citep{Udachin2001} as starting values. Published atom positions 
and lattice parameters \citep{Fortes2007} for Ih and Ic were similarly used. From the refinements,
the lattice parameter, $a$, at each temperature step was obtained and hence the
thermal expansion and density of the cubic clathrate structures
were derived.

\section{Results}

A typical example of a refinement is shown in Fig.~\ref{pawely}, which
shows a comparison of the SXRPD patterns for Ih, Ic and CO$_2$ clathrates
formed in the MS10.5 solution at a CO$_2$ pressure of 10 bar.
The presence of the clathrates at 90~K and 180~K is evident from the formation
of multiple features at 14$^{\circ}-19^{\circ}$, 21.4$^{\circ}$,
23$^{\circ}-24^{\circ}$, 24.6$^{\circ}$ and 25.2$^{\circ}$ 2$\theta$
\citep{Day2015}. 
During fitting the lattice parameters of Ih ice
were initially set to 4.497479{\AA} and 7.322382{\AA} for the $a$
and $c$ axes respectively \citep{Fortes2007}.

Values for the weighted profile ($R_{\rm wp}$) and background-corrected weighted profile 
($R^\prime_{\rm wp}$) fitting agreement parameters between the calculated and experimental 
diffraction data \citep[see][for further details]{young,McCusker1999} --
excluding the ice peaks from the refinement --
were $R_{\rm wp}$ = 6.44\% and 1.21\% and $R_{\rm wp}^\prime = 30.71\%$ and 27.95\%, 
for the structures formed at 90~K and 180~K respectively.
The associated error in the lattice parameter was typically  $\pm 0.001${\AA}. 

The bulk densities of the MgSO$_4$ starting solutions
were measured using a 1000\mic\ PhysioCare concept Eppendorf Reference
pipette to gather a precise volume of solution and weighed using a Mettler Toledo balance
at room temperature.
The solution densities are given in Table~\ref{salts}.

\subsection{Inhibiting effects on clathrate formation}
\label{miller}

\begin{figure}
\begin{center}
\includegraphics[width=9cm,keepaspectratio]{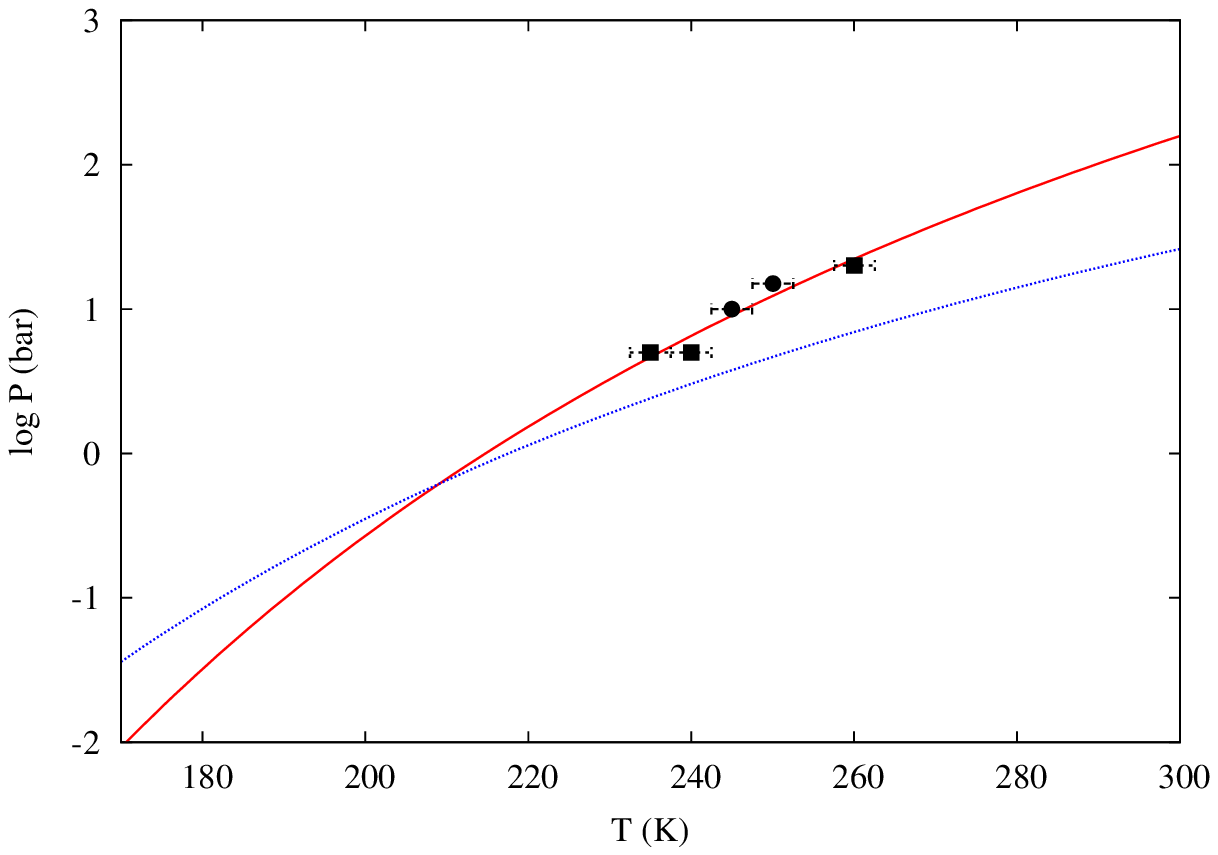}
\caption{CO$_2$ clathrate dissociation curve (red) for CO$_2$ clathrate hydrates in
MS10.5 (circles) and MS3.1 (squares) solutions, compared with the CO$_2$ clathrate dissociation curve
for CO$_2$ clathrate hyrates in pure water \citep[][blue curve]{Miller1961}.
\label{arrenhius}}
\end{center}
\end{figure}

Fig.~\ref{arrenhius} compares the dissociation
temperatures and pressures for the CO$_2$ clathrates formed in the MS10.5
and MS3.1 solutions to those reported by \citet{Miller1961} for pure water.
We have fitted the data for the MS10.5 and MS3.1
solutions, for which we have dissociation temperatures at four pressures
(5, 10, 15, 20~bar), to a function
of the form \citep[cf.][]{Miller1961}
\begin{equation}
 \log_{10} P = -\frac{\alpha}{T} + {\beta} \:\:,
\end{equation}
where $T$ is in~K,  $P$ is in bar, and ${\alpha}$ and ${\beta}$ are constants to be
determined. While we recognise the limited amount of available data
to determine the two parameters $\alpha$ and ${\beta}$,
we find ${\alpha}=1661\pm292$~K and ${\beta}=7.74\pm1.19$.
These values may be compared with those given by \cite{Miller1961} for 
the dissociation of CO$_2$
clathrates in pure water: ${\alpha}=1121.0$~K
and ${\beta}=5.1524$; the data in \citet{Miller1961}
are based on measurements in the temperature range 175--232K. Our data
confirm the likely inhibiting effect by lowering the temperature at which CO$_2$ clathrates
dissociate at a given pressure over the temperature range 235--260~K.

\subsection{Thermal expansion}
\label{TE}
The thermal expansion of clathrate hydrates is an important
property that enables us to understand their physical behaviour.
For example, it has been suggested that the increase in
thermal expansion could be due to greater anharmonicity in
the crystal lattice \citep{Tse1987}; the larger thermal expansion
of clathrates compared to hexagonal
ice could be due to interactions between the guest 
molecule and host structure \citep{Shpakov1998}.

\begin{figure}
\begin{center}
\includegraphics[width=9.cm,keepaspectratio]{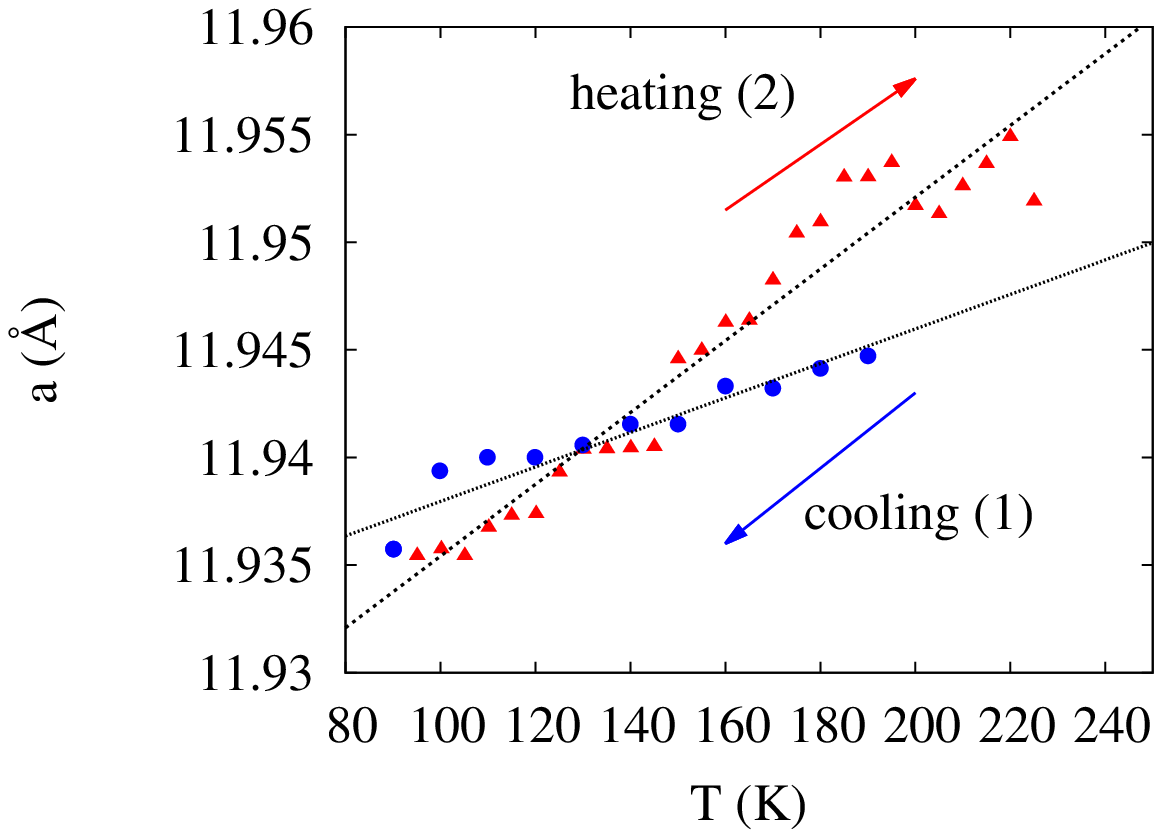}
\includegraphics[width=9.cm,keepaspectratio]{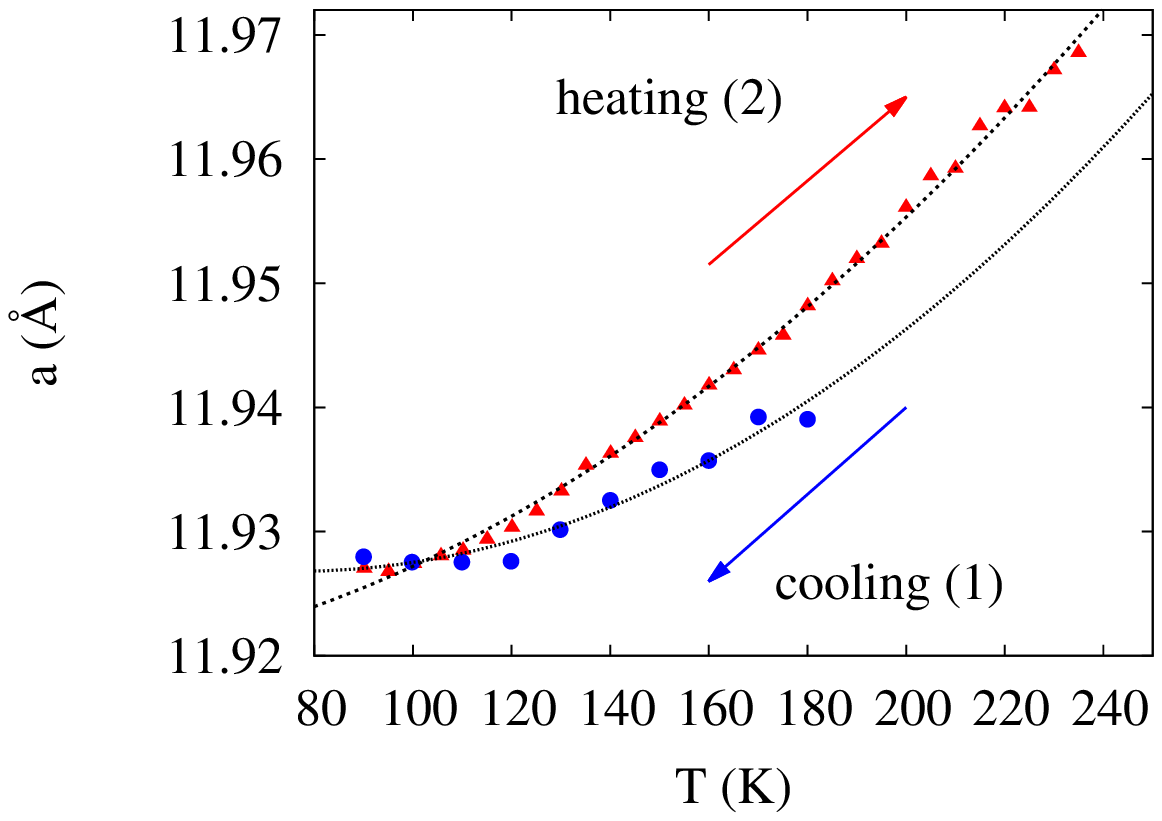}
\includegraphics[width=9.cm,keepaspectratio]{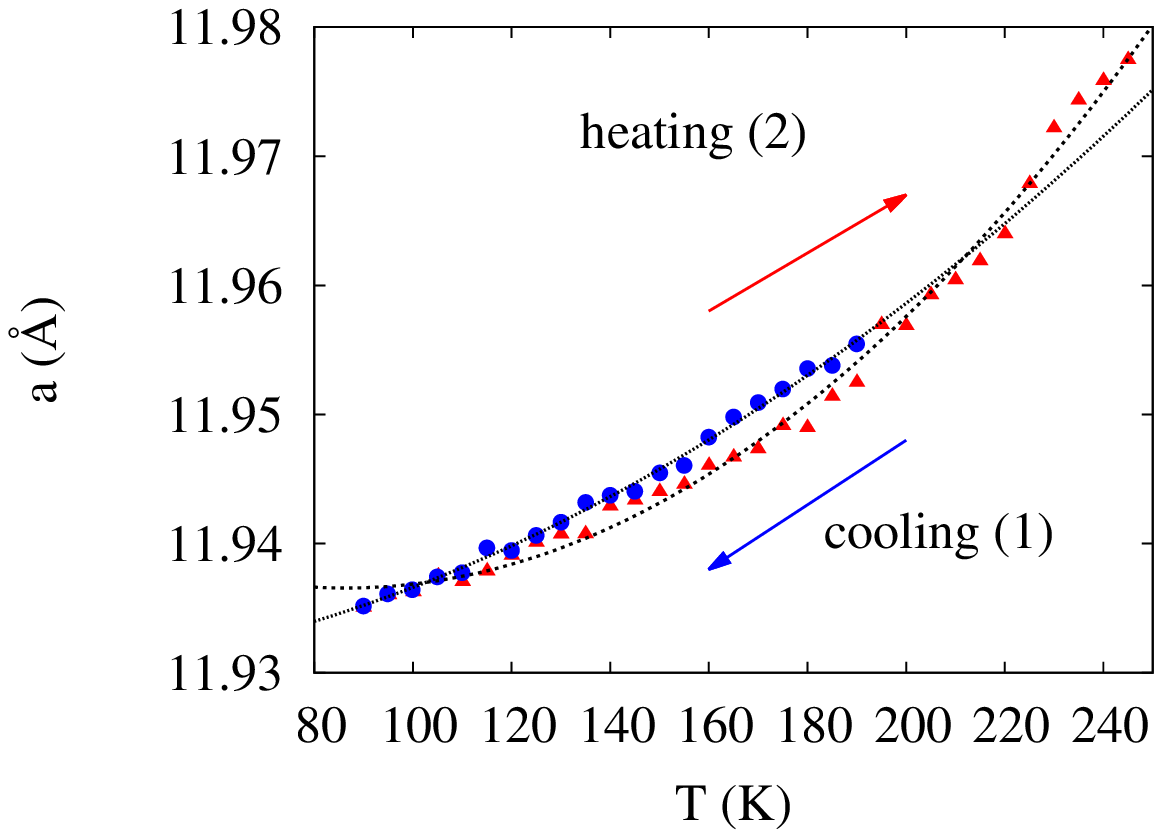}
\caption{The temperature dependence of the lattice parameters of 
CO$_2$ clathrate hydrates formed in solutions of MS10.5 and MS3.1 at various pressures.
From top-bottom:
MS10.5 at 5~bar,
MS10.5 at 10~bar and MS3.1 at 20~bar. 
``(1)'' and ``(2)'' indicate that the cooling was performed first, followed by heating.
For ease of presentation blue symbols represent values obtained
during cooling and red values obtained during heating.
\label{thermalexp}}
\end{center}
\end{figure}

\begin{figure}
\begin{center}
\includegraphics[width=9.4cm,keepaspectratio]{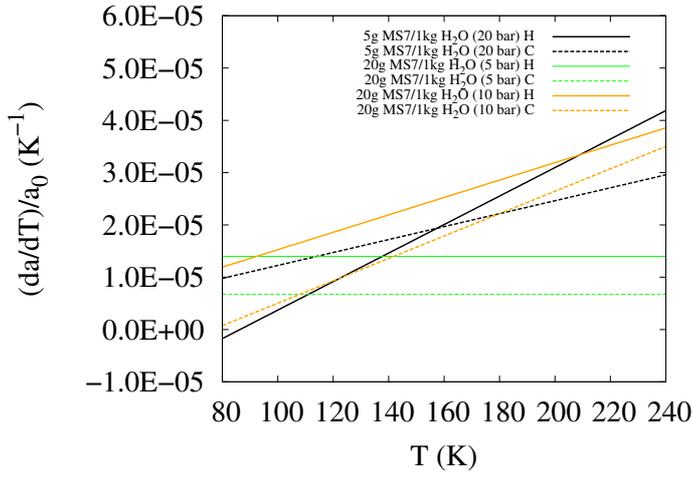}
\caption{The thermal expansion coefficient for CO$_2$ clathrate hydrates formed in 
solutions of MS10.5 and MS3.1 derived from the polynominal fits
to the data in Fig.~\ref{thermalexp}. 
Black: MS3.1/20~bar; green: MS10.5/5~bar; orange: MS10.5/10~bar.
H = Heating and C = Cooling.
\label{alpha}}
\end{center}
\end{figure}

\begin{table*}
\centering
\small
\caption{Coefficients of the polynomial expression for 
describing lattice constants of CO$_2$ clathrate hydrates
formed in the MS10.5 and MS3.1 solutions.
\label{thermalexp-tab}}
\begin{tabular}{c|ccc|ccc}
\hline
 Solution                              &                         & Heating                       &                                &                   & Cooling                      & \\
                               & $a_0$ (\AA)             & $a_1$ ($10^{-5}$\AA\ K$^{-1}$)    & $a_2$  ($10^{-7}$\AA\ K$^{-2}$)    & $a_0$  (\AA)      &  $a_1$  ($10^{-5}$\AA\ K$^{-1}$)  &  $a_2$  ($10^{-7}$\AA\ K$^{-2}$) \\ \hline  
20g MS7/1kg H$_2$O (5 bar)     & 11.9187                 & 16.671                        & ---                            & 11.9299           & $8.01391$        & --- \\
  \{10.5 g MgSO$_4$/kg H$_2$O\}& $\pm 0.001418$          & $\pm 0.8611$                  & ---                            & $\pm 0.001149$    & $\pm 0.8213$     & --- \\  [3ex]
20g MS7/1kg H$_2$O (10 bar)    & 11.9189                 & $-1.64158$                    & $9.91808$                      & 11.9343           & --19.5623        & $12.7815$  \\
  \{10.5 g MgSO$_4$/kg H$_2$O\}& $\pm0.002084$           &  $\pm2.685$                   & $\pm 0.82$                     & $\pm 0.009007$    & $\pm13.77$       & $\pm 5.077$ \\  [3ex]
5g MS7/1kg H$_2$O (20 bar)     & 11.9487                 & --28.1259                     & $16.2818$                      & 11.9294           & 0.144563         & 7.38167 \\
  \{3.1 g MgSO$_4$/kg H$_2$O\} & $\pm0.002782$           & $\pm3.49$                     & $\pm1.034$                     & $\pm 0.002302$    & $\pm 3.395$      & $\pm 1.207$ \\ \hline
\end{tabular}
\end{table*}

Fig.~\ref{thermalexp} shows the dependence of the clathrate
lattice parameter on temperature at three pressure-composition combinations.
We have used a polynomial
approach to describe the temperature dependency of the lattice 
parameter, using the function   
\begin{equation}
 a = a_0 + a_1T + a_2T^2  \:\:.
 \label{TE1}
 \end{equation}
Table~\ref{thermalexp-tab} gives
the values of the coefficients $a_0$, $a_1$, $a_2$, obtained by fitting
Eq.~(\ref{TE1}) to the data.
Although the first order term is apparently not significantly different
from zero in the bottom two rows of the Table, its inclusion 
was found to significantly improve the fit to our experimental data
when inclusion of the second order term is necessary. 

The MS10.5 solution at a CO$_2$ pressure of 10~bar was 
cycled once only (cf. Section~\ref{experimental})
and clathrates
appeared at $185\pm5$~K on cooling from 250~K.
It seems evident from Fig.~\ref{thermalexp}
that the expansion of the clathrate on heating does not follow the behaviour
on cooling: there is hysteresis in that the cooling and heating seem not to be
reversible. The MS10.5 solution at a CO$_2$ pressure of 5~bar
shows a greater degree of hysteresis
compared to the 10~bar solution. It too was thermally cycled
and, on cooling, clathrates appeared at  $195\pm5$~K.
Similarly, the MS3.1 solution at a CO$_2$ pressure of 20~bar
was also cycled once, with clathrates appearing
at $247.5\pm2.5$~K when cooled from 250~K. 
This solution shows a significantly lower degree of hysteresis
compared to the solutions at 5 and 10~bar CO$_2$ pressure.
The difference in behaviour between heating and cooling
may be related to differing levels of bonding disorder 
within the clathrate phase (see discussion in Section~\ref{discussion}).

The coefficient of thermal expansion at constant pressure is defined in
the usual way as $\left[(da/dT)/a_0\right]_P\:$. In the simplest case,
the expansion has a linear dependence on temperature and the
coefficient of expansion is $a_1$, which is independent of temperature.
The coefficients of thermal expansion
are plotted as a function of temperature in Fig.~\ref{alpha},
and exhibit strong pressure dependency. Those CO$_2$
clathrates formed at the lower pressure of 5 bar display a purely linear expansion,
while those formed at higher pressure show more complex behaviour.
Since higher pressures result in higher cage occupancy \citep{Hansen2016},
our results imply that the occupancy of the cages may
influence the thermal expansion of clathrates.
This is discussed further in Section~\ref{discussion}.

\subsection{Density}

\begin{figure}
\begin{center}
\includegraphics[width=9.cm,keepaspectratio]{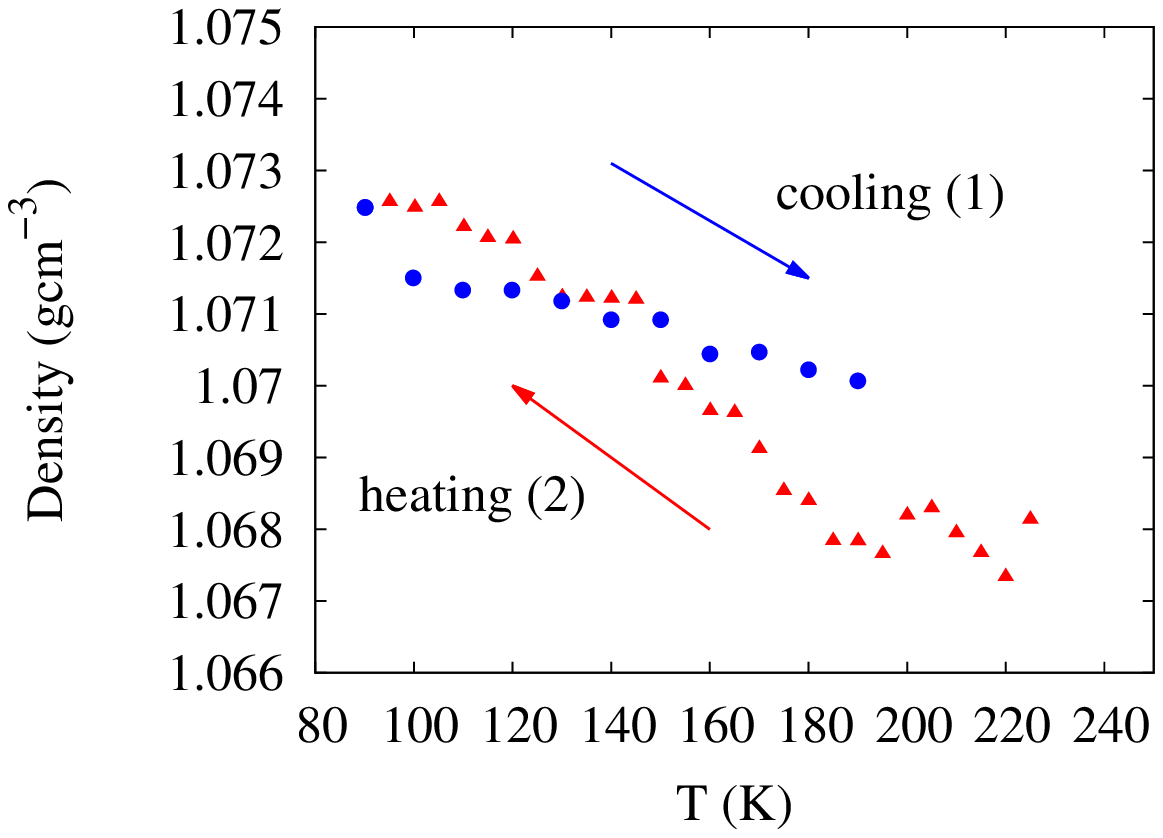}
\includegraphics[width=9.cm,keepaspectratio]{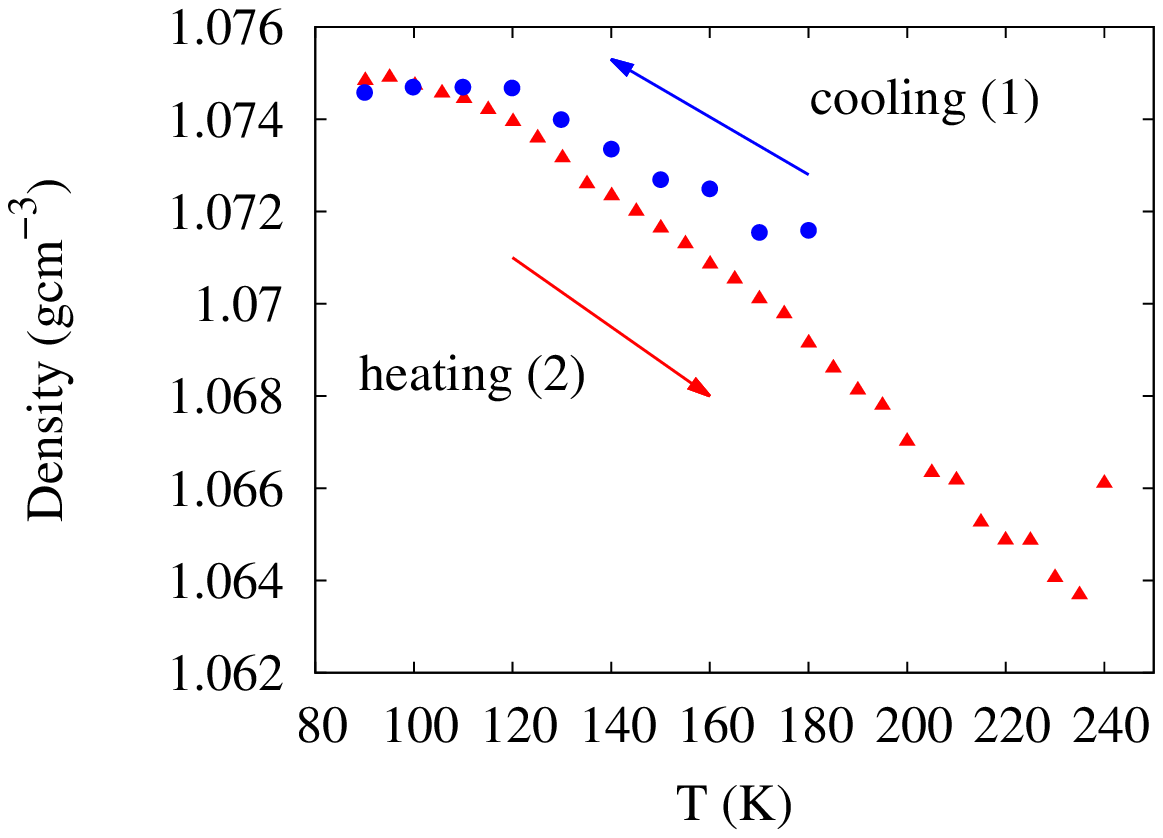}
\includegraphics[width=9.cm,keepaspectratio]{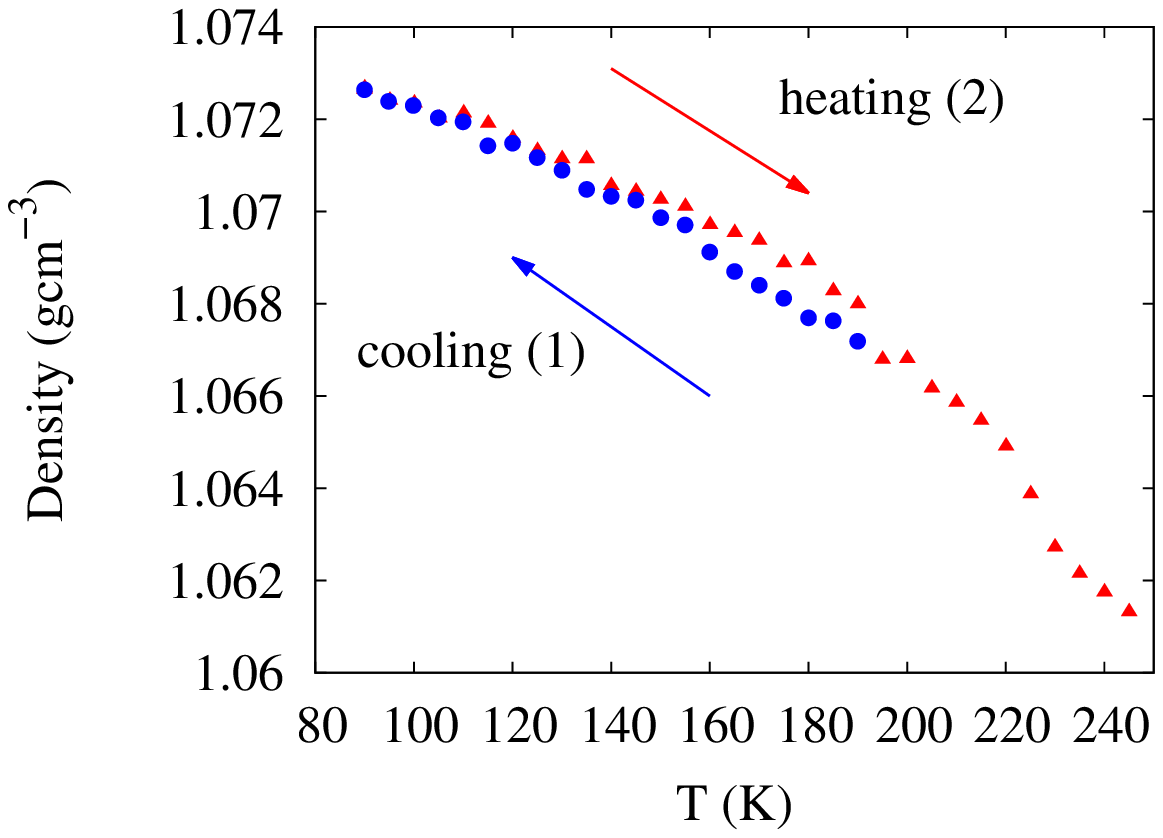}
\caption{Dependence of density on temperature. From top-bottom: MS10.5 at 5~bar,
MS10.5 at 10~bar and MS3.1 at 20~bar. Symbols/colours, 
and meaning of ``(1)'' and ``(2)'',
as per Fig.~\ref{thermalexp}.
\label{density}}
\end{center}
\end{figure}

The density, $\rho$, of a clathrate depends on the lattice parameter, $a$, the mass of its
water molecules, the mass of the guest molecule 
and the cage occupancy; it is calculated as follows \citep{Prieto-Ballesteros2005}:
\begin{equation}
\rho=\frac{\left(M_{\mbox{{\tiny CO$_2$}}} \: (6\theta{_1}+2\theta{_2})+46M_{\mbox{\tiny H$_2$O}}\right)}{a^3}
\:\:.
\label{rho}
\end{equation}
Here M$_{\mbox{{\tiny CO$_2$}}}$ and M$_{\mbox{{\tiny H$_2$O}}}$ are the masses of the CO$_{2}$ guest
 and water molecules respectively, and $\theta{_1}$ and
$\theta{_2}$ are the fractional occupancies of the large and small
cages respectively. Raman data for CO$_2$ clathrates formed at 20 bar in pure water \citep{Day2015}
indicated that only the large clathrate cages are occupied by CO$_2$ \citep[see also][]{Ratcliffe1986}.
For the time being, we assume in the following that this is also the case for clathrates
formed in the presence of MgSO$_4$ and that therefore $\theta_{1}=1$ and $\theta_{2}=0$.
The dependence of the clathrate densities on temperature,
calculated using Eq.~(\ref{rho}), are shown in Fig.~\ref{density}. 
The deduced densities vary with composition and, as would be expected
from the hysteresis effect in the lattice parameter (see Fig.~\ref{thermalexp}), 
on whether the clathrate is being heated or cooled.
We discuss this further in Section~\ref{ddd} below and also consider the effect of
fractional occupancy of the cages.

\subsection{Weight percentage of clathrate, Ih, and Ic ice}
\label{CIhIc}
\begin{figure}
\begin{center}
\includegraphics[width=9cm,keepaspectratio]{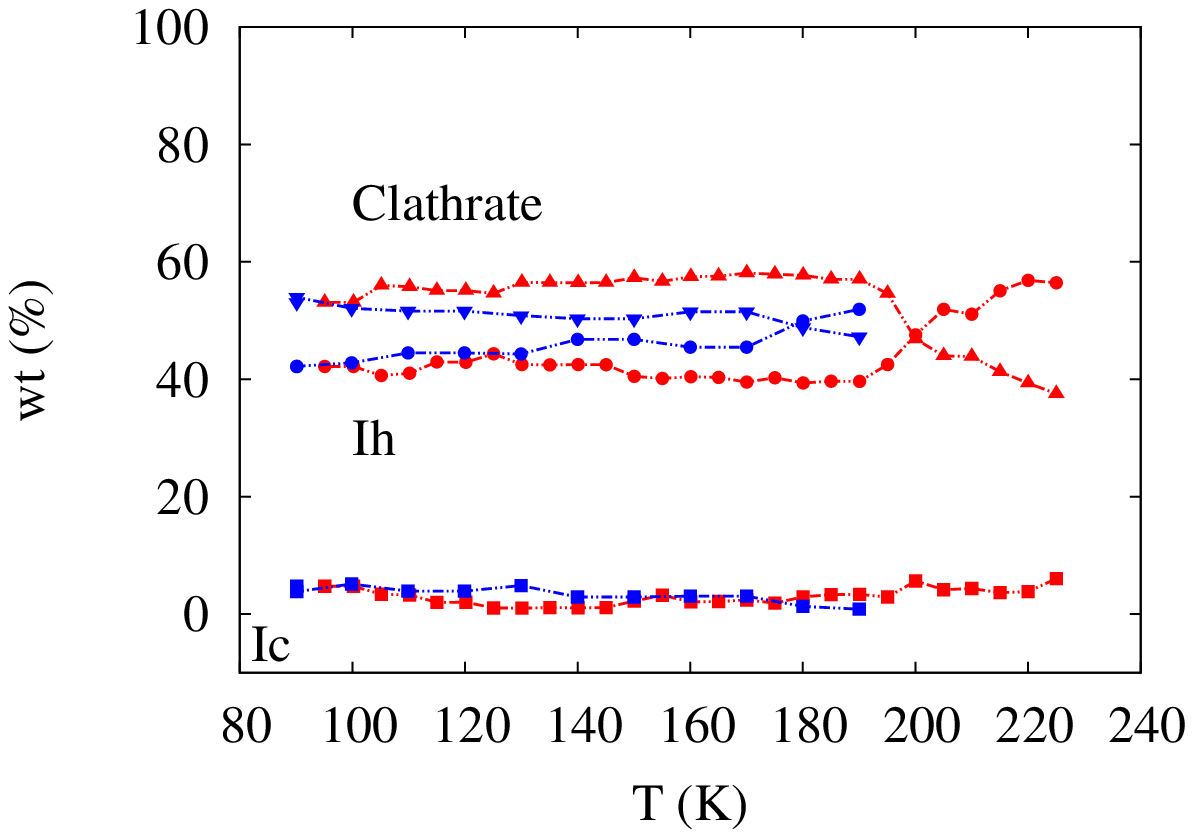}
\includegraphics[width=9cm,keepaspectratio]{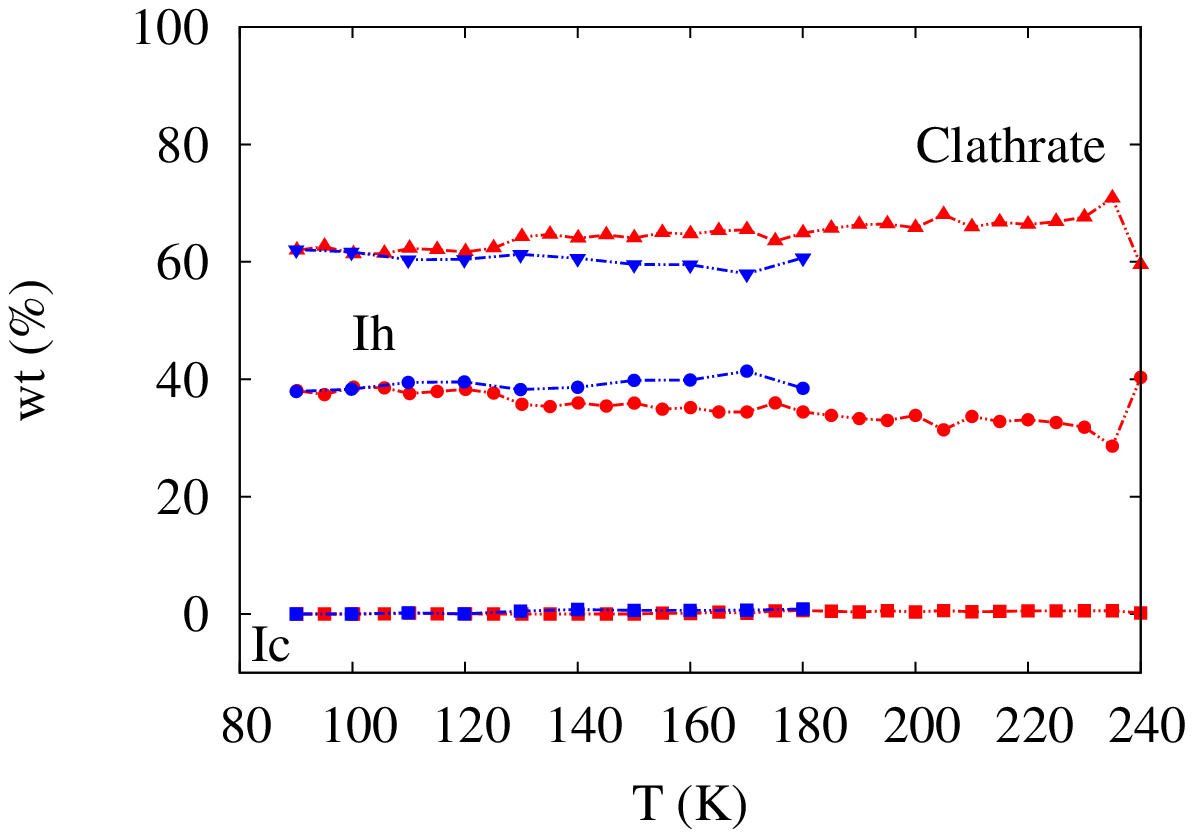}
\includegraphics[width=9cm,keepaspectratio]{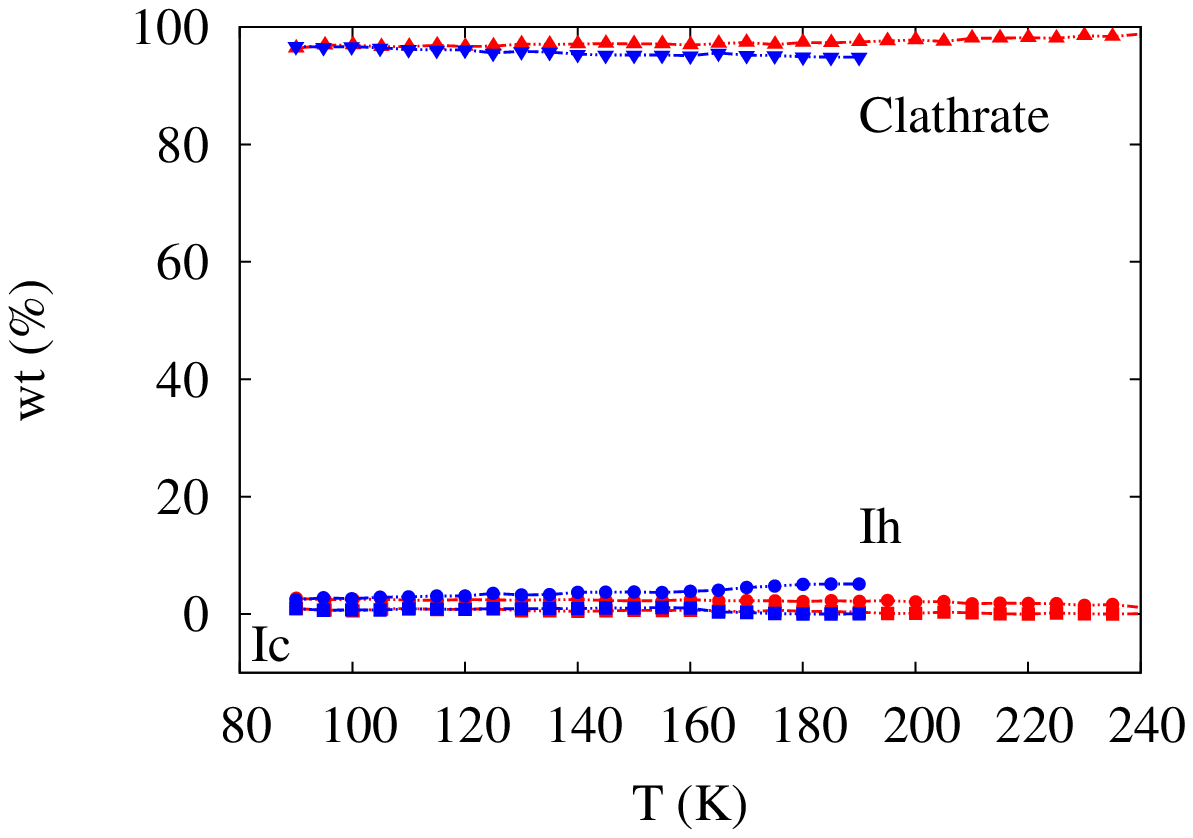}
\caption{Weighted percentage (wt\%) curves for solutions. From top-bottom: MS10.5 at 5~bar,
MS10.5 at 10~bar and MS3.1 at 20~bar. 
Triangles: clathrates; circles: Ih ice; squares Ic ice. Colours are as per Fig.~\ref{thermalexp}.
\label{wt}}
\end{center}
\end{figure}

From the relative contribution of each phase to the overall intensity of 
features in the
powder diffraction pattern we can obtain the relative fraction by weight of each 
crystalline component
present in the sample under study. These are shown in Fig.~\ref{wt} as a function 
of temperature for the MS10.5 (at 5 and 10 bar CO$_2$) and MS3.1 (20 bar CO$_2$).
It is immediately evident that the proportion
of Ic formed in all three samples is small (typically $<5$\%), even at 90~K.

Fig.~\ref{wt} also shows that the relative composition  predominantly depends
on pressure and salt concentration. The MS3.1 solution at a CO$_2$ pressure of 20~bar
is the lowest concentration and highest pressure sample and contains
the highest proportion of clathrates. The other two samples have the same salt
concentration but are at lower pressures (5 and 10~bar) and consequently show 
lower proportions of clathrate. However, in all three samples, the composition of
Ic is similar and always less than 5\%; this is discussed further in
Section~\ref{nature}. 

\section{Discussion}

\label{discussion}

\subsection{Inhibiting effect of MgSO$_4$ on clathrate formation}

It is well known that electrolytes have an inhibiting effect
on the formation of clathrate hydrates \citep{Sabil2009}. This
is caused by the ions in the electrolyte solution
lowering the solubility of the gas, hence lowering
the activity of the water, resulting in the clathrate hydrates
forming at lower temperatures relative to their
development in pure water \citep{Duan2006}. Also, the presence of
inhibitors impedes the water molecules from forming hydrogen
bonds \citep{Sabil2009}, adding a further obstacle to clathrate formation.

Sodium chloride (NaCl) and calcium chloride (CaCl$_{2}$)
electrolyte solutions have been extensively studied as these
are some of the major components of terrestrial seawater and rocks and
their inhibiting effect is well known. They decrease
the dissociation temperature of clathrates by approximately 5~K in
concentrations of $10{\%}$ by weight of solution and by more
than 10~K close to the eutectic solution composition
\citep[17\% by weight;][]{Prieto-Ballesteros2005}.
In situ studies of clathrate hydrates formed in chloride 
solutions will be reported elsewhere (Safi et al., 2017, in preparation).

The best electrolyte inhibitors will exhibit maximum charge and minimum
radius \citep{Makogon1981} and, while less is known about
magnesium electrolyte solutions (e.g. MgCl$_{2}$ and MgSO$_{4}$),
they do exhibit an inhibiting effect
that is stronger than calcium or sodium electrolyte solutions
\citep{Sabil2009}.
The smaller ionic size of magnesium increases the
surface charge density, and so attracts more water molecules, thus
decreasing the activity of water \citep{Sabil2009}.

\citet{Prieto-Ballesteros2005} observed a decrease in the crystallisation 
temperature of clathrates due to the presence of magnesium. However, the
inhibiting effect of the dissolved magnesium in their
experiment was small, amounting to about 2~K at 17\% MgCl$_{2}$.
Also noted by
\citet{Prieto-Ballesteros2005} is that the salt depresses the
freezing point of water by approximately 4~K, and so a larger
temperature difference between ice melting and clathrate dissociation
is observed in the eutectic salt system. A similar trend
was reported earlier by \citet{Kang1998} who found that, as they increased
the concentration of MgCl$_{2}$, the amount of hydrate formed at a
particular pressure becomes less. 

Fig.~\ref{wt} shows that there is a larger difference
in the wt\% of clathrates, during both heating and cooling for
the samples at a CO$_2$ pressure of 10~and 20~bar compared to the samples
at 5~and 10~bar. This could be due to the fact that the
solutions subjected to 5~and 10~bar CO$_2$ contain 20~g of epsomite
per kg water and the solution subjected to a 20~bar CO$_2$ pressure contains 5~g of
epsomite per kg water, i.e. the epsomite is acting as a
clathrate inhibitor. This is further suggested in Fig.~\ref{arrenhius} where we compare our
clathrate dissociation temperatures with the CO$_2$ clathrate dissociation curve given by
\citet{Miller1961}. As discussed in Section~\ref{miller},  this
shows that clathrates formed in the salt solution dissociate
at lower temperatures.

\subsection{Thermal expansion}
\label{TE2}
A surprising feature of the thermal expansion behaviour
(Fig.~\ref{thermalexp}) is the apparent hysteresis
in the dependence of $a$
on $T$, depending on whether the sample is being cooled or
heated. The variation of $a$ with $T$ for the MS3.1 solution at a CO$_2$ pressure of 20~bar
is almost reversible, with little difference
between the cooling and heating curves.
However for the MS10.5 solution at a CO$_2$ pressure of 10~bar we begin
to see a distinct difference between cooling and heating
while for the MS10.5 solution at a CO$_2$ pressure of 5~bar the difference is very noticeable.
The increase in hysteresis with decreasing pressure
may be due to two contributing factors:
\begin{enumerate}
\itemsep=0mm
\item clathrate stability is greater at higher
pressures, so that thermal cycling has a lesser effect;
\item the possibility that
during clathrate formation, the ice-phase water molecules that form
the clathrate cages shift in position and form hydrogen bonds 
with liquid-phase water molecules. The latter  originate from the fluid inclusions/channels
rich in Mg and SO$_4$ ions that result from the eutectic freezing out 
of the pure-phase water ice. This displacement
would cause adjacent water molecules from the surrounding
cages to break hydrogen bonds, hence altering neighbouring
cages \citep{Shin2012}. Such a process would be expected
to affect the elastic properties of the cages, leading to
stress hysteresis \citep{Soh2007}, and to the hysteresis we see in the thermal expansion
(see Fig.\ref{thermalexp}).
This effect should be strongest in those samples
with the highest concentration of epsomite, as is indeed observed.
This may also be related to our observation (see Section~\ref{nature})
that the clathrate structure may play a role in stabilising the
Ih ice phase over the Ic phase.

\end{enumerate}

Furthermore, at higher pressures the cage occupancy 
is higher and could result in an increase of the lattice parameters by 
several thousandth of an \AA\ \citep{Hansen2016}. Indeed, \citeauthor{Hansen2016} 
found that CO$_2$ molecules situated in the small cages expand the clathrate
lattice at higher temperatures. Despite the fact that
we have assumed $\theta_1=1$ and $\theta_2=0$, it may be that the clathrates
formed at higher pressures have a value of $\theta_2>0$.
This is what our experimental data suggest
as the thermal coefficient of expansion for samples at the higher pressure exhibit a steeper gradient
(see Fig~\ref{alpha}).
However it is difficult with the data available to draw any firm conclusions
about the effects of pressure and salinity on the hysteresis in thermal expansion.
Further data, covering a greater area of the pressure and salinity parameter space,
are needed to address this issue.

\subsection{Clathrate density and buoyancy}
\label{ddd}

The variation of clathrate density with temperature
(see Fig.~\ref{density}) has
implications for the sinking or rising capabilities of the
clathrates in planetary oceans. According to our results the general
implication is that the clathrate density is higher at lower
temperatures and lower at higher temperatures, implying
they have a greater probability
of sinking at lower temperatures and of floating at higher temperatures.
However, this will also depend on the salinity of the ocean
in question.

The MS3.1 and MS10.5 solutions used in this experiment 
are similar to
the salinities of the oceans on Enceladus and Europa
(see Table~\ref{salts}).
If we compare our clathrate densities 
at various temperatures with that of the solutions in which they were formed, 
we see that both the measured solution densities (1.003~g~cm$^{-3}$ and 
1.016~g~cm$^{-3}$ for MS3.1 and MS10.5 solutions, respectively) 
are much lower than the CO$_2$ clathrate density (Fig.~\ref{density}), 
irrespective of the temperature and pressure 
conditions. The higher clathrate densities suggest the clathrates would always sink.
Indeed, this is also true
if we assume $\theta_2$ is between 0.625--0.688 (keeping $\theta_1=1$)
which are the values \citet{Hansen2016} obtained from their experimental
investigation. Therefore sinking of clathrates is the likely scenario
for both Enceladus and Europa. 

For CO$_2$ clathrates to float in an ocean with salinity
close to Enceladus' and Europa's
they would need lower guest molecule occupancy. From Eq.~(\ref{rho})
the clathrates formed in the MS3.1 solution at a CO$_2$
pressure of 20~bar would require the larger
cages to be 73\% filled, while the MS10.5 solutions at CO$_2$ pressures of 5 and 10~bar
would need the larger cages to be 78\% and 77\% filled respectively (assuming $\theta_2=0$). 
As mentioned,  pressure directly affects the clathrate cage
occupancy in that occupancy
(and hence density) is greater at increased pressure. 
A consequence of this is that  clathrates 
formed deeper in an ocean would have a higher occupancy and
would therefore have a greater probability of sinking.

Our results also suggest that if CO$_2$ clathrates were to form
at the base of a floating ice shell \citep[cf.][]{Prieto-Ballesteros2005},
they too should sink and transport encased gases
to the bottom of the ocean floor. If, on the other hand,
the ocean was of eutectic composition of MgSO$_4$ \citep[17~wt\%, as
suggested for Europa by][]{Prieto-Ballesteros2005},
the ocean density would be
1.19~g~cm$^{-3}$, implying the CO$_2$ clathrates would in fact float; this
would cause fracturing and gravitational collapse of the terrain
due to rapid release of gas \citep{Prieto-Ballesteros2005}.

The effect of pressure on density can be seen in Fig.~\ref{density}. If we compare
both the MS10.5 solutions at CO$_2$ pressures of 5 and 10~bar, we see that
the sample subjected to 5~bar CO$_2$ pressure has a lower density from
95~K to 195~K on heating. From 195~K onwards the sample subjected to a
CO$_2$ pressure of 10~bar produces the lowest density clathrates.
However, considering the cooling curves in Fig.~\ref{density}, the MS10.5 
solution at a CO$_2$ pressure of 10~bar has the highest density throughout the entire
cooling process.  For most of the
heating curves and all of the cooling curves in Fig.~\ref{density} our results are 
consistent with the conclusion that higher pressure environments
produce clathrates with higher densities and which are therefore less buoyant.
We should note that our conclusion regarding buoyancy relates
to the case of CO$_2$ clathrates.
For the case of multiple-guest occupancy (e.g. CO$_2$ + CH$_4$) 
the sinking/floating capabilities might well be different. However such
multiple-guest clathrates would most likely be of the less common sH type
(see Section~\ref{ChSS}).

\subsection{The nature of the ice}
\label{nature}
Ic is the most common polymorph
of ice at  temperatures below 160~K. Therefore investigation
into the nature of the ice phase that coexists with clathrates is especially significant 
in the context of cold extra-terrestrial environments as it would impact 
on the interpretation of remotely sensed data and our understanding of the physical processing
and conditions in these environments \citep{falenty09}.
Above 240~K water crystallises into the thermodynamically favoured Ih phase, the rate
of change of Ih to Ic being highest between 200~K -- 190~K, while
below $\sim160$~K Ic is the stable phase
\citep{Falenty2015}.
We note that the experimental work of \citeauthor{Falenty2015} was undertaken 
at 6~mbar, typical of Mars' atmosphere; however the crystallisation temperatures
of Ih and Ic do not seem to be sensitive to pressures up to a few 100~bar
\citep[see e.g.][]{zhang2015}.

Despite being thermodynamically favourable at low temperature, our data
show that the wt\% of Ic is never more than
about 5\%, even at 90~K (see Fig.~\ref{wt}) and it may be that
the clathrate structure preferentially stabilises the Ih phase over the Ic phase.
However, it is also possible that, at low temperatures, the rate
of transformation is slow and, given sufficient time,
all of the Ih would eventually transform to Ic.
Furthermore, Ic might be the more favourable phase if the ice were to condense
at temperatures lower than those used in this work.

Also, we note that clathrate dissociation does not 
occur until the temperature reaches approximately 200--240\,K, at which point 
ice would form as Ih after 
the clathrates have dissociated \citep{falenty09}. 
Therefore, since the CO$_2$ clathrates formed at the the 
pressures and salinities used in this experiment does not seem to dissociate 
at very low temperatures below 240~K, we would not expect the formation of
significant quantities of the Ic phase.

\section{Conclusion}

By the use of in situ synchrotron X-ray powder diffraction we have
studied the formation, dissociation and thermal expansion properties 
of CO$_2$ clathrate hydrates formed in MgSO$_4$ salt solutions.
Specifically, we have:
\begin{enumerate}
\itemsep=0mm
\item found the dissociation temperatures and pressures of CO$_2$
clathrate hydrates formed in a salt solution containing epsomite
(MgSO$_{4}\cdot$7H$_2$O) and that the salt
solution inhibits clathrate formation. 
\item
computed the density of these clathrates at different temperatures
and pressures and compared this to the density of the solution in which they
were formed. While it was found that the density of the clathrate depends on temperature 
and pressure (and hence local factors such as seasonal and tidal changes), 
when compared to the density of the salt solution they formed in they 
should in general sink, irrespective of the temperature and pressure.
\item
investigated the polymorphs
of the associated ice phases. We report the dominance of Ih throughout
the experiment despite the expectation of Ic being
the thermodynamically stable polymorph at lower temperatures.
This may be due to the salt solution inhibiting the Ih to Ic transformation. However
further investigation into the thermodynamics and kinetics
of ice in relation to clathrates is needed to confirm this.
\end{enumerate}
These experimental observations demonstrate the importance
of understanding the role played by salts, clathrates and ice on the surface geology
and sub-surface oceans of icy Solar System bodies. 
As a gas transport mechanism the likely sinking of CO$_2$ clathrates formed in saline environments could
make a significant contribution to  ocean floor geochemistry on such objects.

\begin{acknowledgements}
We thank an anonymous referee for their careful and thorough reading of the paper, and for
making several comments and suggestions that have helped to clarify and improve the text.
This work was supported by the Diamond Light Source through beamtime awards EE-9703 and EE-11174.
ES is supported by Keele University and Diamond Light Source.
\end{acknowledgements}

\end{document}